\documentclass[12pt,preprint]{aastex}
\usepackage{epstopdf}
\usepackage{chngpage}
\usepackage{graphicx}
\usepackage{natbib}

\shorttitle{IMPACTS BETWEEN ASTEROIDS AND NSS}
\shortauthors{Geng \& Huang}

\begin{document}

\title{FAST RADIO BURSTS: COLLISIONS BETWEEN NEUTRON STARS AND ASTEROIDS/COMETS}

\author{J. J. Geng\altaffilmark{1, 2}, and Y. F. Huang\altaffilmark{1, 2}}

\altaffiltext{1}{School of Astronomy and Space Science, Nanjing University, Nanjing 210046, China; hyf@nju.edu.cn}
\altaffiltext{2}{Key Laboratory of Modern Astronomy and Astrophysics (Nanjing University), Ministry of Education, Nanjing 210046, China}

\begin{abstract}
Fast radio bursts (FRBs) are newly discovered radio transient sources. Their high dispersion measures
indicate an extragalactic origin. But due to the lack of observational data in other wavelengths,
their progenitors still remain unclear. Here we suggest the collisions between neutron stars
and asteroids/comets as a promising mechanism for FRBs. During the impact process, a hot plasma fireball
will form after the material of the small body penetrates into the neutron star surface.
The ionized matter inside the fireball will then expand along the magnetic field lines.
Coherent radiation from the thin shell at the top of the fireball will account for the observed
FRBs. Our scenario can reasonably explain the main features of FRBs, such as their durations,
luminosities, and the event rate. We argue that for a single neutron star, FRBs are not likely
to happen repeatedly in a forseeable time span since such impacts are of low probability.
We predict that faint remnant X-ray emissions should be associated with FRBs,
but it may be too faint to be detected by detectors at work.
\end{abstract}


\section{INTRODUCTION}
\label{sect:intro}

Recently, the discovery of a number of fast radio bursts (FRBs) was reported \citep{Lorimer07,Keane11,Thornton13,Burke14,Spitler14,Ravi15}.
Typically, they are single radio pulses with flux densities $S_{\nu} \sim \rm{a~few}~\rm{Jy}$ and durations
$\delta t \sim \rm{a~few}~\rm{ms}$ at frequency $\nu_{\rm FRB} \sim 1~\rm{GHz}$. No counterparts in other
wavelengths have been detected yet \citep{Petroff15}, maybe due to the lack of rapid, multiwavelength follow-up after the bursts. According to their high dispersion measures ($\sim 500 - 1000~\rm{cm}^{-3}~\rm{pc}$),
FRBs may originate at cosmological distances \citep{Thornton13}, although the possibility that they happened in the dense regions of local galaxies still cannot be excluded yet \citep{Katz14b,Luan14,Pen15}. If FRBs are at cosmological distances with redshifts $0.5 \leq z \leq 1$ \citep{Thornton13}, the characteristic isotropic radio luminosity ($L_{\rm FRB}$) will be $\sim 10^{42-43}~\rm{erg}~\rm{s}^{-1}$ and the isotropic energy ($E_{\rm FRB}$) released is then $10^{39-40}~\rm{erg}$.
The observable event rate of FRBs is suggested to be $\sim 10^{4}~\rm{sky}^{-1}~\rm{day}^{-1}$ \citep{Thornton13,Keane15}.

The central engines of FRBs are under hot debate. The durations of FRBs indicate the
emission regions are compact, while the high brightness of the radio emission requires coherent emission
to take effect \citep{Katz14a,Luan14}, which is similar to radio emissions from pulsar magnetospheres \citep{Ruderman75,Cheng77,Benford77}. Also, the energy reservoir in a neutron star (NS) magnetosphere
is high enough to account for the energy release of FRBs. These few but significant clues motivate some authors
to associate FRBs with scenarios involving NSs. Several possible models have been proposed, e.g.,
magnetar giant flares \citep{Popov07,Kulkarni14,Lyubarsky14,Pen15}, collapse of hypermassive neutron stars (NSs)
into black holes \citep{Falcke14,Zhang14,Ravi14}, binary NS mergers \citep{Totani13} and planetary companions around NSs \citep{Mottez14}. At the same time, other kinds of models have also been proposed, e.g., binary white dwarf mergers \citep{Kashiyama13}, flare stars \citep{Loeb14} and evaporation of primordial black holes \citep{Barrau14}.

However, besides the above scenarios,
there may be another external way to trigger the energy release in the NS magnetosphere.
Pulsar timing observations
have revealed that there may be planets \citep{Wolszczan92} or asteroid belts \citep{Shannon13}
around NSs. Recent study shows transient, supergiant pulses from active or dormant pulsars may be triggered by
debris entering their magnetospheres \citep{Cordes15}. By migrating into the NS light cylinder
before completely destroyed by evaporation and ionizing, the debris can also disrupt current flows and electromagnetic radiation and therefore account for some intermittency seen in pulsars \citep{Cordes08}.
On the other hand, it has previously been suggested in the literature that small solid bodies such as
asteroids or comets can impact NSs occasionally \citep{Colgate81,Mitrofanov90,Katz94,Huang14}.
By referring to impact, we mean we are considering small bodies of ballistic trajectories
(with low angular momentum), as apposed to orbiting objects with significant angular momentum.
It is interesting to notice that the timescale and energy release
of these impacts can meet the requirements of FRB progenitors from the first view.
Thus we suggest that the impacts between
NSs and comets/asteroids may provide a possible explanation for FRBs.

The structure of our paper is as follows. We briefly describe the impact process in Section 2. In Section 3,
we present the formulas for the radiation process and derive the basic features of the emission.
In Section 4, the remnant X-ray emission in our scenario is discussed.
Our conclusions are summarized in Section 5.

\section{IMPACT PROCESS}
\label{sect:impact}
In our modeling, the calculations are mainly based on the assumption that FRBs are at
cosmological distances with redshifts $0.5 \leq z \leq 1$.
The absence of correlation of FRBs
with any known galaxies or galaxy clusters indicates a lower bound on their distance
($\sim$ 100 Mpc, \cite{Katz14b}). Note that if FRBs happen at such a local distance,
our model can still work. In that case,
we will only need a smaller asteroid to collide with the NS.
A more detailed discussion on this point is given in the last section of our paper.

Direct impacts between asteroids/comets and NSs have been previously
discussed in different contexts. In order to give a quantitative
description of our model, here we would like to give a brief review of the
impact process, mainly following the study in \cite{Colgate81}.
When a small solid body of mass $m$ falls freely in the gravitational field
of an isolated NS of mass $M$, it will undergo elongation in the radial direction.
The elongated body will be broken up at the breakup radius and the collapsed material will
be compressed by gravitational radial acceleration and magnetic fields of the NS before landing.
For an Fe-Ni asteroid with a density $\rho_0$, radius $r_0$, and shear strength $s$,
the breakup radius of the elongated body is
\begin{equation}
R_{b} = \left(\rho_0 r_0^2 M G / s \right)^{1/3},
\end{equation}
where $G$ is the gravitational constant. We assume the leading fragment
(at $R_b-r_0$) and lagging portion (at $R_b+r_0$) have the same velocity $v_b$
($v_b$ is determined from the free fall assumption from $R = \infty$)
when the asteroid center is at $R_b$. The subsequent free fall
gives the evolution of the velocities of the leading and lagging fragments ($v_-$ and $v_+$) as
\begin{equation}
\frac{1}{v_{\pm}} \approx \left(\frac{2 G M}{R}\right)^{-1/2} \left(1 \pm \frac{r_0 R}{2 R_{b}^{2}}\right).
\end{equation}
The difference of arrival time at the surface of the NS ($R_{\rm{NS}}$) is then
\begin{eqnarray}
\Delta t_a = \int_{R_{\rm NS}}^{R_{b}+r_0} \frac{\rm{d} R}{v_{+}} - \int_{R_{\rm NS}}^{R_{b}-r_0} \frac{\rm{d} R}{v_{-}} \simeq 2 r_0 / v_{b} = 2 r_0 \left(\frac{2 G M}{R_{b}}\right)^{-1/2} \nonumber\\
= 1.58 \times 10^{-3} m_{18}^{4/9} s_{10}^{-1/6} \left(\frac{\rho_0}{8~\rm{g~cm}^{-3}}\right)^{-5/18} \left(\frac{M}{1.4 M_{\odot}}\right)^{-1/3} \rm{s},
\end{eqnarray}
where the convention $Q_x = Q/10^x$ in cgs units is adopted hereafter.
This impact time scale is less than the duration of the observed FRBs,
so the short-time characteristic of FRBs can basically be met in our model.

The descriptions above only consider the gravitational influence of the NS
on a potential impacting body. In realistic case, as the asteroid moves toward the NS,
it may be evaporated and ionized by radiation from
the compact star and its magnetosphere before impacting the NS \citep{Cordes08}.
Note that in our model, the conditions are somewhat special.
First, the asteroids are of low angular momentum and possess ballistic trajectories,
as opposed to orbiting objects that are spiraling in.
Second, the mass of the asteroid is relatively large (typically $\sim 10^{18}$ g, see Section 3 below).
It is much larger than the characteristic mass discussed by \cite{Cordes08} for which
the effects of evaporation and ionization are important.
Third, the asteroid is assumed to be of Fe-Ni composition and the shear strength will be larger.
As a result, the asteroid might be less prominently affected by factors
other than the gravitational influence. We now present a detailed discussion
on this point.

For slowly spinning old pulsars, ionization does not happen until the
asteroid has entered the magnetosphere.
The luminosity of X-rays from the magnetosphere may not be high
if the tilt angel of the dipolar magnetic field is small \citep{Cordes08}
so that we only need to consider the radiation from the NS surface here.
For an old NS with a surface
temperature of $T_{\rm NS}$, assuming that the asteroid at $R$ is in thermal equilibrium with the
NS surface radiation, then the temperature of the asteroid is
$T_{\ast} = T_{\rm NS} \left(R_{\rm NS} / 2 R\right)^{1/2} \simeq 707~T_{\rm{NS},5} R_{10}^{-1/2}$ K.
When $T_{\ast}$ reaches the iron evaporation point ($\simeq$ 2000 K),
the corresponding distance is smaller than typical $R_{b}$ ($\simeq 2 \times 10^9$ cm).
It means the evaporation process could be ignored before the tidal breakup.
We now consider possible electrodynamic effects imposed by the magnetosphere.
After the accreted matter enters the magnetosphere, it will be conducting because
of the immense electric field in the frame of the matter. The magnetic skin depth will
be small ($\sim 10^{-2}$ cm) since the conductivity is large \citep{Colgate81}. Therefore,
when the accretion column penetrates between two surfaces of constant longitude,
it can be treated as a diamagnetic body with all field lines parallel to its surface.
On the other hand, if the accretion column crosses some regions with a nonzero field-aligned
electric field ($E_{\parallel} \neq 0$, e.g., \cite{Takata06}),
particle acceleration in these regions can yield $\gamma$-ray emission
that drives electron/positron pair cascades.
With these free charges,
$E_{\parallel}$ of these regions may actually turn to be zero since the magnetosphere
tends to become force-free. As a result, in our model, $E_{\parallel}$ may not have a strong
influence on the trajectory of the main part of the incoming matter
at the expense of a little front intruding matter.
Thus the result of Equation (3) will not be seriously affected
after considering the effects of the NS magnetic field in our framework.

When approaching the NS surface, the accretion column will
penetrate the magnetic field as a compressed sheet of diamagnetic fluid
with all magnetic field lines parallel to its surface.
The compression in longitude reduces the thickness of the sheet to a few millimeters,
while its width in latitude would expand to a few kilometers at the NS surface.
The dense matter then plunges into the NS outer crust and the kinetic energy
is converted to thermal energy, launching a rapidly expanding plasmoid fireball
along the field lines (see Figure 1 for a schematic illustration). A fan of field lines
is finally filled with
hot plasma \citep{Tademaru71,Colgate81}. In the plasma, plenty of
electrons are accelerated to ultra-relativistic speeds by magnetic reconnection near the collision site.
Radiation from this fan of hot plasma will give birth to an observable FRB, as
detailed in the following section.

\section{EMISSION MECHANISM}
\label{sect:emission}
The high brightness temperatures of FRBs indicate they are connected
with coherent emission. This radiation mechanism may also be
involved in radio emission of pulsars.
Although the progenitors of FRBs are still uncertain,
some constraints on the emission region can be derived from observations
\citep{Katz14b}. In our scenario, the hot plasma fan can produce the required
coherent emission (see Figure 1). The electron bunches originated from the
collision will form a shell with a thickness of
$\Delta$ at $r_{\rm emi}$ from the NS. The duration $\delta t$ of FRBs implies
$\Delta \approx c \delta t$. The emission volume of this shell is
$V_{\rm emi} \approx 4 \pi f \Delta r_{\rm emi}^2$.
Note that $f$ is the ratio of the shell solid angle to $4 \pi$.
At the bottom of the fan, the thickness in longitude is $\sim 10^4$ cm
as a result of the turbulent expansion, thus a rough estimate gives
$f \sim \frac{10^4~\rm{cm}}{2 \pi R_{\rm NS}} = 3 \times 10^{-3}$.
Electrons radiate coherently in patches with a characteristic radial size of
$\lambda = c/\nu_{\rm c}$ ($\nu_{\rm c}$
is the characteristic frequency of curvature emission), the corresponding volume of each patch is
$V_{\rm coh} = \left(4/\gamma^2\right) r_{\rm emi}^2 \times \left(c/\nu_{\rm c}\right)$.
Here the factor $4/\gamma^2$ is the solid angle within which electrons
can be casually connected in the relativistic beam.

The coherent curvature emission luminosity can be estimated as \citep{Kashiyama13}
\begin{equation}
L_{\rm tot} \approx \left(P_{\rm e} N_{\rm coh}^{2}\right) \times N_{\rm pat},
\end{equation}
where $P_{\rm e} = 2 \gamma^4 e^2 c / 3 r_{\rm emi}^2$ is the emission power of a single electron,
$N_{\rm coh} \approx n_{\rm e} \times V_{\rm coh}$ is the number of electrons in each
coherent patch, and $N_{\rm pat} \approx V_{\rm emi}/V_{\rm coh}$ is the number of the patches.
The characteristic frequency of curvature emission is
\begin{equation}
\nu_{c} = \gamma^3 \frac{3 c}{4 \pi r_{\rm emi}}.
\end{equation}

On the other hand, the coherent radiation mechanism may be effective only within a certain
distance $r_{\rm max}$ from the pulsar, since the filamentary instability would grow beyond
$r_{\rm max}$ \citep{Benford77}. Above $r_{\rm max}$, the transverse pressure of
plasma begins to exceed the magnetic energy density and the coherent emission vanishes \citep{Benford77}.
As $r_{\rm emi}$ may be slightly less than $r_{\rm max}$, we introduce a parameter
$\epsilon$ to describe the deviation from the balance between the plasma pressure
and the magnetic energy density, i.e. $n_{\rm e} \gamma m_{\rm e} c^2 = \epsilon B^2(r_{\rm emi}) / 8 \pi$.
Here the magnetic field strength can be estimated as
$B(r_{\rm emi}) \approx B_{\rm NS} \times \left(r_{\rm emi}/R_{\rm NS}\right)^{-3}$,
where $B_{\rm NS}$ is the surface magnetic field strength at $R_{\rm NS}$.

Using the formulas above and assuming $L_{\rm tot} = f L_{\rm FRB}$, $\nu_{c} = \nu_{\rm FRB}$,
we can solve out the typical Lorentz factor of electrons in the emitting shell as
\begin{equation}
\gamma \simeq 547 \left(\epsilon_0^{2} B_{\rm{NS},12}^4 R_{\rm{NS},6}^{12} \delta t_{-2}
\nu_{\rm{FRB},9}^9 L_{\rm{FRB},42}^{-1}\right)^{1/30}.
\end{equation}
The corresponding typical values of other quantities are
$r_{\rm emi} \simeq 1.2 \times 10^9$ cm, $\Delta \simeq 3 \times 10^8$ cm,
$B(r_{\rm emi}) \simeq 580$ G, and $n_{\rm e} \simeq 3 \times 10^7$ cm$^{-3}$.
It is worthy to note that $n_{\rm e}$ actually refers to the
number density of electrons and positrons generated from photon pair production.
Therefore, it could be significantly larger than the Goldreich-Julian density \citep{Goldreich69}.
Also, it can be found that $\gamma$ and other quantities are insensitive to $f$
(or the real emission volume).
Actually, the emission volume can be larger than the volume derived from
$f \sim 3 \times 10^{-3}$ above.

For the radio emission propagating through the plasma, it is essential
that the characteristic plasma frequency must be below the frequency of the propagating radio waves, i.e.,
\begin{equation}
\nu_{\rm p} = \gamma \left(\frac{n_{\rm e}^{\prime} e^2}{\pi m_e}\right)^{1/2} \leq \nu_{\rm FRB},
\end{equation}
where $n_{\rm e}^{\prime} = n_{\rm e}/\gamma$ is the number
density of electrons in the comoving frame.
Using the parameters derived above, we find this requirement can be satisfied.

In general, both the gravitational potential energy of the asteroid
and the magnetic field energy of the NS can provide the energy emitted.
If all the energy released is contributed by the former one, i.e.,
\begin{equation}
f E_{\rm FRB} = \eta_{\rm R} \frac{G M m}{R_{\rm NS}},
\end{equation}
then the mass needed is
$m = 5.4 \times 10^{17} \eta_{\rm{R},-2}^{-1} f_{-3} E_{\rm{FRB},40} R_{\rm{NS},6} M_{1.4 M_{\odot}}^{-1}$ g,
where $\eta_{\rm R}$ is the efficiency of transforming the potential energy into radio radiation
and we adopt $\eta_{\rm R} \sim 10^{-2}$ as the typical value in the following calculations.
This mass is roughly in the mass range of normal asteroids, assuring the self-consistency of our model.

\section{X-RAY EMISSION}
\label{sect:X-ray}
No counterparts associated with FRBs have been observed at wavelengths
other than the radio range till now, making FRBs more mysterious.
A recent multiwavelength follow-up to FRB 140514 reveals no
variable counterparts or transient emissions associated with it \citep{Petroff15}.
There are two possible reasons for this mysterious fact.
On one hand, the progenitors of FRBs may be such special transient
sources so that the duration of signals in other bands are also too short,
beyond the reaction capability of telescopes in search.
On the other hand, the flux of FRB counterparts may be very weak and be below the detection
limit of the telescopes. In our scenario, the matter
collapsed onto the NS surface may contribute to emissions
in other bands. It is interesting to discuss whether the
remnant emissions can be detected by detectors at work.

While some electrons are accelerated to move to the top of the
fan, most of the matter collapsed would remain in a column on
the NS surface. The temperature of this matter is high during the
impact, and would decrease later when it loses its energy by radiation.
Although the cooling process of the hot matter is not clearly known,
we can have a rough estimate on the basis of reasonable assumptions.
After the giant flare of 1998 August 27, transient X-ray emission
decaying as $\propto t^{-0.7}$ was observed from SGR 1900+14.
It was suggested to be the cooling behavior of the heated crust
of magnetar \citep{Lyubarsky02}.
In our model, we assume that the radiation from the heated matter
is thermal and the cooling obeys the same decaying law.
Thus, for a FRB at a luminosity distance of $d_{\rm L}$,
the remnant X-ray light curve is
\begin{equation}
F_{\rm X} \approx \sigma T^4 \left(\frac{R_{\rm NS}}{d_{\rm L}} \right)^2,
\end{equation}
where $\sigma$ is the Stefan-Boltzmann constant and $T$ is the temperature of the matter.
The $t^{-0.7}$ decaying law indicates that the matter cools as $T \propto t^{-7/40}$.
We further assume the ratio of the energy released in X-ray band
to the potential energy is $\eta_{\rm X}$.
Then we can obtain $F_{\rm X}$ and compare it with the sensitivity of current detectors.
Figure 2 illustrates the X-ray light curves calculated using different $\eta_{\rm X}$
(ranging from $10^{-2}$ to 1.0),
together with the sensitivity line of the $\it{Swift}$/X-Ray Telescope \citep{Moretti09,Burrows14,Yi14}.
From this figure, it can be seen that the remnant X-ray emission after
the collision is well below the sensitivity line and cannot be detected.

Other factors, e.g., the spreading of the collapsed matter on the NS surface,
and the rotation of the NS, would reduce the flux significantly
and are not considered here. The calculation above is actually an optimistic estimate.
Therefore, the X-ray counterparts (not in the bursting phase) associated with
FRBs may not be detected by detectors at work.

\section{SUMMARY AND DISCUSSION}
\label{sect:disc}
In this study, we propose that the impacts between NSs and asteroids/comets
may be a promising mechanism for FRBs.
For an asteroid of typical mass of $10^{18}$ g falling onto the NS surface,
a fan of hot plasma would form after the millisecond collision.
The consequent emitting shell at $r_{\rm emi} \sim 1.2 \times 10^9$ cm,
containing electrons/positrons with $\gamma \approx 550$,
will emit in radio wavelength coherently. The main characteristics of FRBs,
including the timescale and luminosity, can be well explained in our scenario.
However, the remnant X-ray emission following the FRBs will be faint
according to the calculations under reasonable assumptions.

In our scenario, $\gamma$-ray photons may be emitted, e.g.,
by the inverse Compton scattering of transient energetic
electron/positron pairs on the soft photons within burst timescale.
The $\gamma$-ray event may be as short as the radio burst itself,
but might last for a rotation period or longer.
The fraction of electron/positron pairs with a high Lorentz factor
to trigger $\gamma$-ray emission is highly uncertain,
but we can give a rough estimate based on some simplifications.
We assume the efficiency of transferring the potential energy into
$\gamma$-ray emission is $\eta_{\gamma}$ and
the duration $\tau$ is the same to that of
radio burst, i.e., $\tau \sim \delta t \simeq 1$ ms.
Then we can find the average $\gamma$-ray flux during $\tau$ as
$\sim \frac{\eta_{\gamma}}{\eta_{\rm R}} \frac{f E_{\rm FRB}}{4 \pi d_{\rm L}^2 \tau}
= 10^{-19} f_{-3} \eta_{\gamma,-2} S_{\nu,1 \rm{Jy}} \nu_{\rm{FRB},9} \tau_{-3}^{-1}~\rm{erg}~\rm{cm}^{-2}~\rm{s}^{-1}$,
which is too low to detect.

The main calculations in this paper are based on
the assumption that the FRBs are generated at cosmological distances.
Nevertheless, our scenario can also work for FRBs that somehow happen in local galaxies.
For an FRB happens at $d_{\rm L} \sim$ 100 Mpc, its isotropic radio energy
is then $\sim 10^{37}$ erg. Using Equation (8), we can find that
the corresponding asteroid mass needed is only $\sim 5 \times 10^{15}$ g.
Although the distance is much smaller in this case,
the remnant X-ray flux is still low since the potential energy
(and thus the X-ray energy release) is accordingly reduced.
However, note that the dynamics of small objects ($\sim 10^{15}$ g) could be significantly
different from that of large objects ($\sim 10^{18}$ g), because
the evaporation and electrodynamic effects may be stronger \citep{Cordes08}.
A detailed study on these effects will be helpful but it is beyond the scope
of this paper.

Till now, no repeating bursts have been observed from any
particular FRB sources. This feature can be reasonably explained
in our scenario since a direct collision between an
asteroid/comet and the NS can only trigger the burst once,
and such collisions are not likely to happen repeatedly on
short timescales. These predictions make our model testable
by more observations in the future.

The event rate of FRBs is another crucial clue to the progenitors.
The very limited field of view of current large radio
telescope makes the observation of large numbers of FRBs very difficult.
Although only a few FRBs have been observed till now,
it is widely believed that the actual event rate should
be very high, i.e., $\sim 10^{4}~\rm{sky}^{-1}~\rm{day}^{-1}$ \citep{Keane15}.
We need to know whether this event rate could be explained in our framework.
Although it is difficult to assess the exact event rate of
our impact processes theoretically, we can give a rough estimate on
the basis of some reasonable assumptions.
Several mechanisms have been proposed for the formation of
planetary systems around NSs. The planetary system can be formed
along with the progenitor \citep{Wolszczan92,Phinney93,Podsiadlowski93},
from the fall back material after the supernova \citep{Lin91},
or from the material provided by a lower mass companion \citep{Nakamura91,Shannon13}.
For a NS with a planetary system, it is possible that
some asteroids would impact the NS. First, asteroids could be gravitationally
disturbed by other planets and be scattered toward the NS \citep{Guillochon11}.
Second, the planets may have chances to collide with each other
and produce clumps falling to the NS \citep{Katz94}.
In previous studies, the recurrence time ($\tau_{\rm rec}$)
of strong direct impacts (with $m \geq 10^{18}$ g) in a
typical NS planetary system
is estimated to be $\sim 10^6 - 10^7$ years \citep{Tremaine86,Mitrofanov90,Katz94,Litwin01}.
Noting that the co-moving volume of $z \leq 1$ contains $\sim 10^9$ late-type galaxies \citep{Thornton13},
and supposing the number of NSs in a typical galaxy is $10^8$ \citep{Timmes96},
the observable event rate of the impact is
$\zeta \approx f \frac{10^9 \rm{galaxies} \times 10^8 / \rm{galaxy}}{\tau_{\rm rec}} \simeq 10^4 - 10^5  f_{-3}~\rm{sky}^{-1}~\rm{day}^{-1}$.
Recently the Alfv\'en wing structures formed
during the interaction between a relativistic pulsar wind and
the orbiting small body were investigated. It was found that the Alfv\'en wing
structures will lead the small bodies in a retrograde orbit to
move toward the central NS more rapidly \citep{Mottez11},
which can further increase the collision rate.
Considering these, the theoretical event rate can well
explain the observed FRB rate.

By now, what we are confident about is that the radio emission of FRB
should be coherent given the millisecond duration and a lower bound on the distance.
In fact, what we have proposed is one kind of process that can trigger
the coherent emission from the magnetosphere of NS.
The conditions needed to drive the particle bunching,
which may involve some kinds of instabilities,
is still unclear and is beyond the scope of our work.
Future advance in the study of coherent radiation could give
more rigorous clues to the progenitors of FRBs.

\begin{figure}
   \begin{center}
   \includegraphics[width=\linewidth]{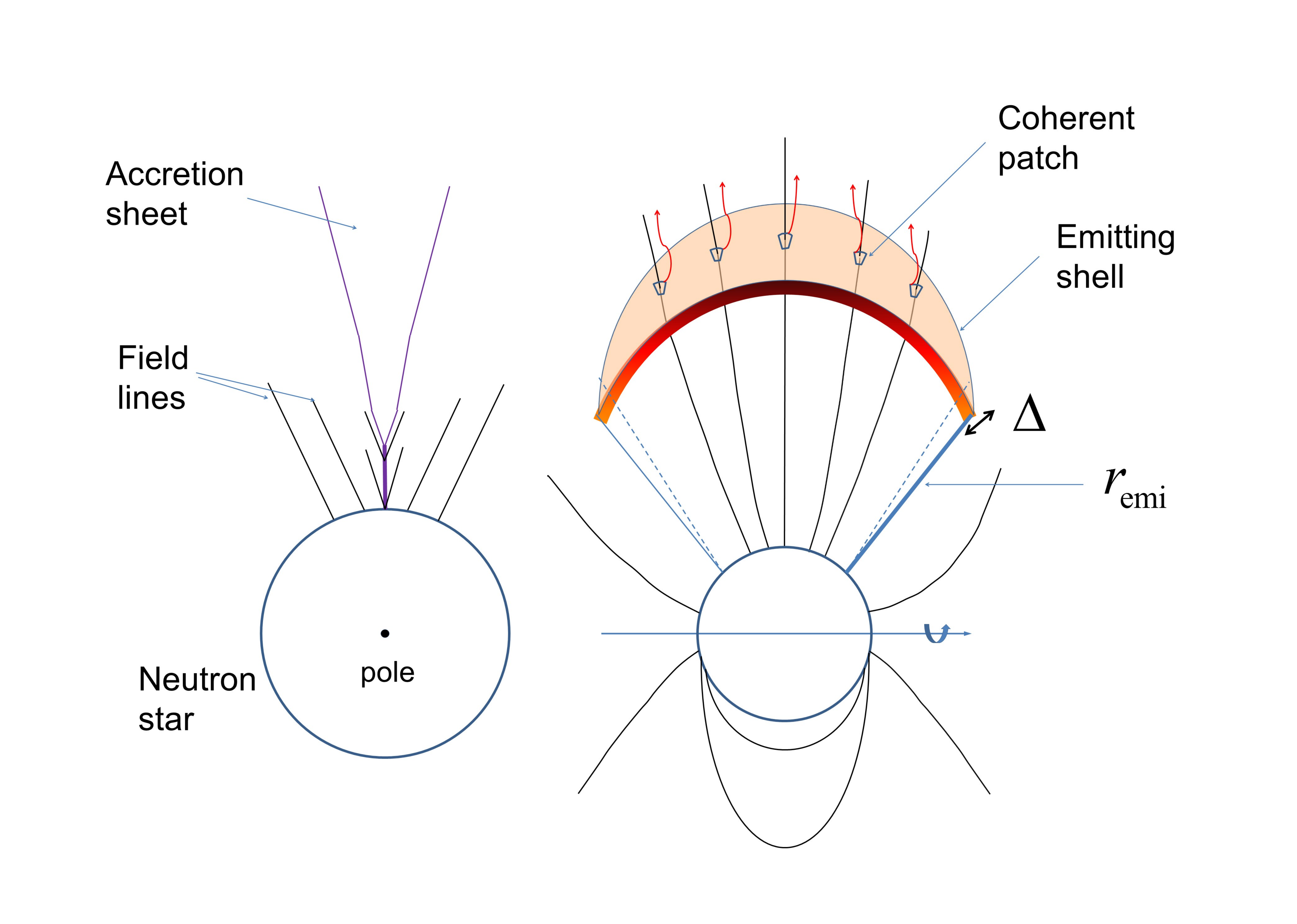}
   \caption{Schematic illustration of the impact between a NS and an asteroid/comet.
   The left panel shows the elongated body is accreted onto the NS as a sheet (see \cite{Colgate81} for a detailed
   plot). The magnetic compression in longitude reduces its thickness to a few millimeters, while the width in latitude would expand to a few kilometers when approaching the NS surface. The right panel depicts the hot plasma fan generated shortly after the collision. The emitting shell (red-orange arc) is the region where the coherent curvature radiation is generated.}
   \end{center}
\end{figure}

\begin{figure}
   \begin{center}
   \includegraphics[width=\linewidth]{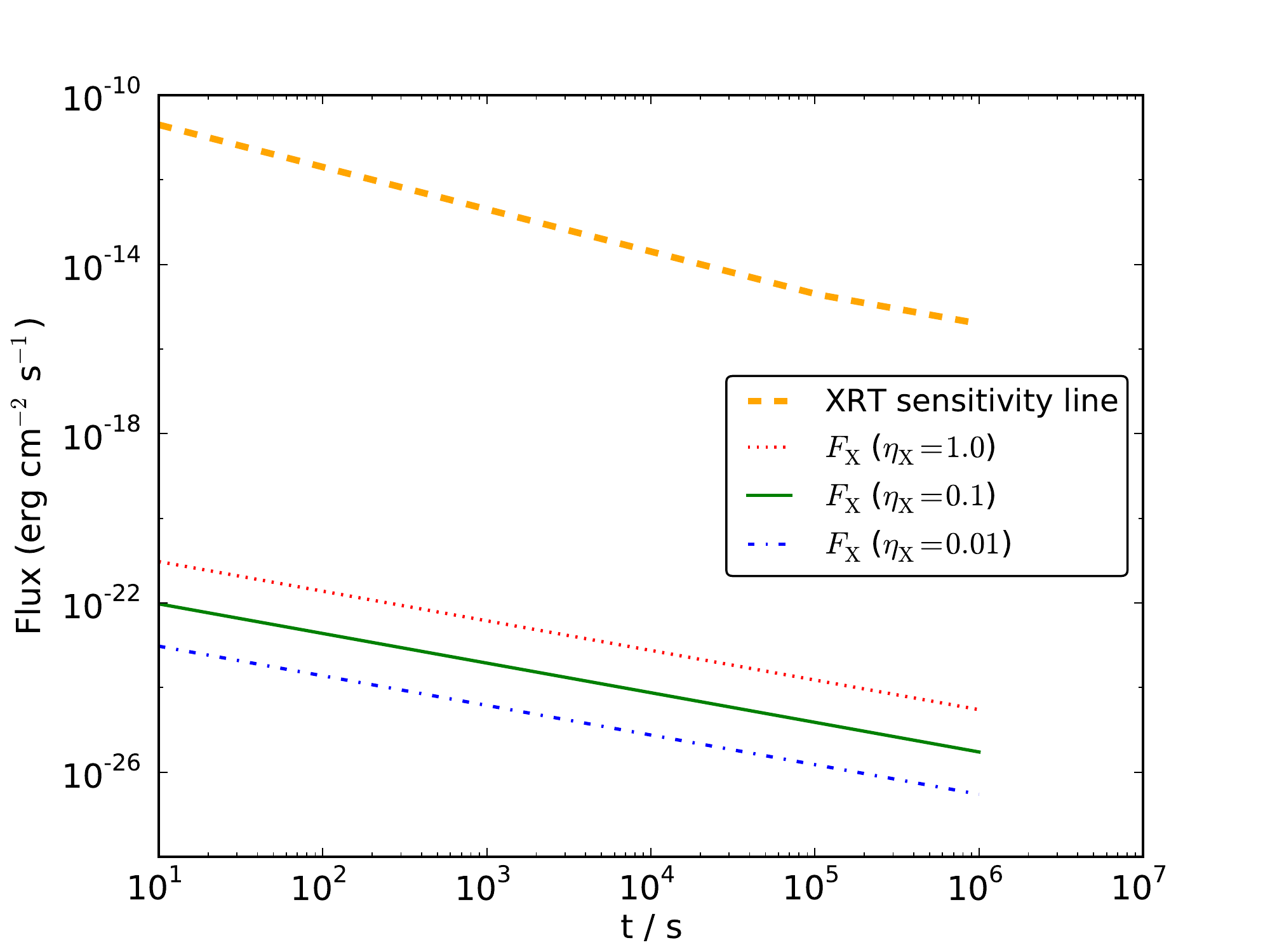}
   \caption{Predicted remnant X-ray light curves after the FRBs in our scenario.
   Three values of $\eta_{\rm X} = 1.0$ (dotted line), 0.1 (solid line), 0.01
   (dash-dotted line) are adopted during the calculation using Equation (9).
   Other quantities are set as typical values, i.e. $\eta_{\rm R} = 0.01$,
   $f = 10^{-3}$, $E_{\rm FRB} = 10^{40}$ erg.
   The thick dashed line is the sensitivity line of the $\it{Swift}$/X-Ray Telescope \citep{Moretti09,Burrows14,Yi14}.
   Note that for the sensitivity line, the X-axis is the integration time, while the Y-axis is the corresponding sensitivity limit under
   this exposure time.}
   \end{center}
\end{figure}

\begin{acknowledgments}
We appreciate many helpful comments and suggestions from the anonymous referee.
We thank Shuang-Xi Yi and Wei Su for helpful discussions.
This study was supported by the National Basic Research Program of China with Grant No.
2014CB845800, and by the National Natural Science Foundation of China with Grant No. 11473012.
\end{acknowledgments}


\begin{thebibliography}{99}
\bibitem[Barrau et al.(2014)]{Barrau14} Barrau, A., Rovelli, C.,
\& Vidotto, F.\ 2014, \prd, 90, 127503

\bibitem[Benford
\& Buschauer(1977)]{Benford77} Benford, G., \& Buschauer, R.\ 1977, \mnras, 179, 189

\bibitem[Burke-Spolaor
\& Bannister(2014)]{Burke14} Burke-Spolaor, S., \& Bannister, K.~W.\ 2014, \apj, 792, 19

\bibitem[Burrows et al.(2014)]{Burrows14} Burrows, D.~N., Frank,
K.~A.,
\& Park, S.\ 2014, BAAS, 223, 353.14

\bibitem[Cheng
\& Ruderman(1977)]{Cheng77} Cheng, A.~F., \& Ruderman, M.~A.\ 1977, \apj, 212, 800

\bibitem[Colgate
\& Petschek(1981)]{Colgate81} Colgate, S.~A., \& Petschek, A.~G.\ 1981, \apj, 248, 771

\bibitem[Cordes
\& Shannon(2008)]{Cordes08} Cordes, J.~M., \& Shannon, R.~M.\ 2008, \apj, 682, 1152

\bibitem[Cordes
\& Wasserman(2015)]{Cordes15} Cordes, J.~M., \& Wasserman, I.\ 2015, arXiv:1501.00753

\bibitem[Falcke
\& Rezzolla(2014)]{Falcke14} Falcke, H., \& Rezzolla, L.\ 2014, \aap, 562, A137

\bibitem[Goldreich
\& Julian(1969)]{Goldreich69} Goldreich, P., \& Julian, W.~H.\ 1969, \apj, 157, 869

\bibitem[Guillochon et al.(2011)]{Guillochon11} Guillochon, J.,
Ramirez-Ruiz, E., \& Lin, D.\ 2011, \apj, 732, 74

\bibitem[Huang
\& Geng(2014)]{Huang14} Huang, Y.~F., \& Geng, J.~J.\ 2014, ApJL, 782, L20

\bibitem[Kashiyama et al.(2013)]{Kashiyama13} Kashiyama, K., Ioka,
K., \& M{\'e}sz{\'a}ros, P.\ 2013, ApJL, 776, L39

\bibitem[Katz et al.(1994)]{Katz94} Katz, J.~I., Toole, H.~A.,
\& Unruh, S.~H.\ 1994, \apj, 437, 727

\bibitem[Katz(2014a)]{Katz14a} Katz, J.~I.\ 2014a, \prd, 89,
103009

\bibitem[Katz(2014b)]{Katz14b} Katz, J.~I.\ 2014b,
arXiv:1409.5766

\bibitem[Keane et al.(2011)]{Keane11} Keane, E.~F., Kramer, M.,
Lyne, A.~G., Stappers, B.~W., \& McLaughlin, M.~A.\ 2011, \mnras, 415, 3065

\bibitem[Keane
\& Petroff(2015)]{Keane15} Keane, E.~F., \& Petroff, E.\ 2015, \mnras, 447, 2852

\bibitem[Kulkarni et al.(2014)]{Kulkarni14} Kulkarni, S.~R., Ofek,
E.~O., Neill, J.~D., Zheng, Z., \& Juric, M.\ 2014, \apj, 797, 70

\bibitem[Lin et al.(1991)]{Lin91} Lin, D.~N.~C., Woosley,
S.~E., \& Bodenheimer, P.~H.\ 1991, \nat, 353, 827

\bibitem[Litwin
\& Rosner(2001)]{Litwin01} Litwin, C., \& Rosner, R.\ 2001, Physical Review Letters, 86, 4745

\bibitem[Loeb et al.(2014)]{Loeb14} Loeb, A., Shvartzvald, Y.,
\& Maoz, D.\ 2014, \mnras, 439, L46

\bibitem[Lorimer et al.(2007)]{Lorimer07} Lorimer, D.~R., Bailes,
M., McLaughlin, M.~A., Narkevic, D.~J.,
\& Crawford, F.\ 2007, Science, 318, 777

\bibitem[Luan
\& Goldreich(2014)]{Luan14} Luan, J., \& Goldreich, P.\ 2014, ApJL, 785, L26

\bibitem[Lyubarsky(2014)]{Lyubarsky14} Lyubarsky, Y.\ 2014, \mnras,
442, L9

\bibitem[Lyubarsky et al.(2002)]{Lyubarsky02} Lyubarsky, Y.,
Eichler, D., \& Thompson, C.\ 2002, ApJL, 580, L69

\bibitem[Mitrofanov
\& Sagdeev(1990)]{Mitrofanov90} Mitrofanov, I.~G., \& Sagdeev, R.~Z.\ 1990, \nat, 344, 313

\bibitem[Moretti et
al.(2009)]{Moretti09} Moretti, A., Pagani, C., Cusumano, G., et al.\ 2009, \aap, 493, 501

\bibitem[Mottez
\& Heyvaerts(2011)]{Mottez11} Mottez, F., \& Heyvaerts, J.\ 2011, \aap, 532, A22

\bibitem[Mottez
\& Zarka(2014)]{Mottez14} Mottez, F., \& Zarka, P.\ 2014, \aap, 569, A86

\bibitem[Nakamura
\& Piran(1991)]{Nakamura91} Nakamura, T., \& Piran, T.\ 1991, ApJL, 382, L81

\bibitem[Pen
\& Connor(2015)]{Pen15} Pen, U.-L., \& Connor, L.\ 2015, arXiv:1501.01341

\bibitem[Petroff et al.(2015)]{Petroff15} Petroff, E., Bailes,
M., Barr, E.~D., et al.\ 2015, \mnras, 447, 246

\bibitem[Phinney
\& Hansen(1993)]{Phinney93} Phinney, E.~S., \& Hansen, B.~M.~S.\ 1993, in ASP Conf. Ser. 36, Planets Around
Pulsars, ed. J.~A. Phillips, S.~E. Thorsett, \& S.~R. Kulkarni (San Francisco,
CA: ASP), 371

\bibitem[Podsiadlowski(1993)]{Podsiadlowski93} Podsiadlowski, P.\ 1993,
in ASP Conf. Ser. 36, Planets Around Pulsars, ed. J.~A.
Phillips, S.~E. Thorsett, \& S.~R. Kulkarni (San Francisco, CA: ASP), 149

\bibitem[Popov
\& Postnov(2007)]{Popov07} Popov, S.~B., \& Postnov, K.~A.\ 2007, arXiv:0710.2006

\bibitem[Ravi
\& Lasky(2014)]{Ravi14} Ravi, V., \& Lasky, P.~D.\ 2014, MNRAS, 441, 2433

\bibitem[Ravi et al.(2015)]{Ravi15} Ravi, V., Shannon, R.~M.,
\& Jameson, A.\ 2015, ApJL, 799, L5

\bibitem[Ruderman
\& Sutherland(1975)]{Ruderman75} Ruderman, M.~A., \& Sutherland, P.~G.\ 1975, \apj, 196, 51

\bibitem[Shannon et al.(2013)]{Shannon13} Shannon, R.~M., Cordes,
J.~M., Metcalfe, T.~S., et al.\ 2013, \apj, 766, 5

\bibitem[Spitler et al.(2014)]{Spitler14} Spitler, L.~G., Cordes,
J.~M., Hessels, J.~W.~T., et al.\ 2014, \apj, 790, 101

\bibitem[Tademaru(1971)]{Tademaru71} Tademaru, E.\ 1971, \apss, 12, 193

\bibitem[Takata et al.(2006)]{Takata06} Takata, J., Shibata, S.,
Hirotani, K., \& Chang, H.-K.\ 2006, \mnras, 366, 1310

\bibitem[Thornton et al.(2013)]{Thornton13} Thornton, D.,
Stappers, B., Bailes, M., et al.\ 2013, Science, 341, 53

\bibitem[Timmes et al.(1996)]{Timmes96} Timmes, F.~X., Woosley,
S.~E., \& Weaver, T.~A.\ 1996, \apj, 457, 834

\bibitem[Totani(2013)]{Totani13} Totani, T.\ 2013, \pasj, 65, L12

\bibitem[Tremaine
\& Zytkow(1986)]{Tremaine86} Tremaine, S., \& Zytkow, A.~N.\ 1986, \apj, 301, 155

\bibitem[Wolszczan
\& Frail(1992)]{Wolszczan92} Wolszczan, A., \& Frail, D.~A.\ 1992, \nat, 355, 145

\bibitem[Yi et al.(2014)]{Yi14} Yi, S.-X., Gao, H.,
\& Zhang, B.\ 2014, ApJL, 792, L21

\bibitem[Zhang(2014)]{Zhang14} Zhang, B.\ 2014, ApJL, 780, L21

\end{thebibliography}
\end{document}